\documentclass
[floats,superscriptaddress,balancelastpage,prl,a4paper,twocolumn,tightenlines,showpacs,10pt]{revtex4}%
\usepackage{graphicx}
\usepackage{amsmath}
\usepackage{amsfonts}
\usepackage{amssymb}%
\begin{document}
\title{Experimental demonstration of a non-destructive controlled-NOT quantum gate
for two independent photon-qubits}
\author{Zhi Zhao}
\affiliation{Department of Modern Physics, University of Science
and Technology of China, Hefei, Anhui 230027, People's Republic of
China} \affiliation{Physikaliches Institut, Universit\"{a}t
Heidelberg, Philisophenweg 12, 69120 Heidelberg, Germany}
\date{\today }
\author{An-Ning Zhang}
\affiliation{Department of Modern Physics, University of Science
and Technology of China, Hefei, Anhui 230027, People's Republic of
China}
\author{Yu-Ao Chen}
\affiliation{Department of Modern Physics, University of Science
and Technology of China, Hefei, Anhui 230027, People's Republic of
China}
\author{Han Zhang}
\affiliation{Department of Modern Physics, University of Science
and Technology of China, Hefei, Anhui 230027, People's Republic of
China}
\author{Jiang-Feng Du}
\affiliation{Department of Modern Physics, University of Science
and Technology of China, Hefei, Anhui 230027, People's Republic of
China}
\author{Tao Yang}
\affiliation{Department of Modern Physics, University of Science
and Technology of China, Hefei, Anhui 230027, People's Republic of
China}
\author{Jian-Wei Pan}
\affiliation{Department of Modern Physics, University of Science
and Technology of China, Hefei, Anhui 230027, People's Republic of
China} \affiliation{Physikaliches Institut, Universit\"{a}t
Heidelberg, Philisophenweg 12, 69120 Heidelberg, Germany}
\date{\today }

\begin{abstract}
Universal logic gates for two quantum bits (qubits) form an
essential ingredient of quantum information processing. However,
the photons, one of the best candidates for qubits, suffer from
the lack of strong nonlinear coupling required for quantum logic
operations. Here we show how this drawback can be overcome by
reporting a proof-of-principle experimental demonstration of a
non-destructive controlled-NOT (CNOT) gate for two independent
photons using only linear optical elements in conjunction with
single-photon sources and conditional dynamics. Moreover, we have
exploited the CNOT gate to discriminate all the four Bell-states
in a teleportation experiment.

\end{abstract}

 \maketitle

The controlled-NOT (CNOT) or similar logic operations between two
individual quantum bits (qubits) are essential for various quantum
information protocols such as quantum communication
\cite{bennett92,bennett93,gisin02} and quantum computation
\cite{shor97}. In recent years, certain quantum logic gates have
been experimentally demonstrated, for example, in ion-traps
\cite{blatt03,wineland03} and high-finesse microwave cavities
\cite{haroche99}. These achievements open many possibilities for
future quantum information processing (QIP) with single atoms.
Another promising system for QIP is to use single photons. This is
due to the photonic robustness against decoherence and the
availability of single-qubit operation. However, it has been very
difficult to achieve the necessary logic operations for two
individual photon-qubits since the physical interaction between
photons is much too small.

Surprisingly, Knill, Laflamme and Milburn (KLM) has shown that
nondeterministic quantum logic operations can be performed using
linear optical elements, additional photons (ancilla) and
postselection based on the output of single-photon detectors
\cite{klm01}. The original proposal by KLM, though elegant, is not
economical in its use of optical components. Various schemes have
been proposed to reduce the complexity of the KLM scheme while
improve its theoretical efficiency
\cite{koashi01,pittman01,ralph02}. Remarkably, a recent scheme
proposed by Nielsen \cite{nielsen04} suggests that without using
the elaborate teleportation and Z-measurement error correction in
the KLM scheme, any non-trivial linear optical gate that succeeds
with finite probability is sufficient to obtain efficient quantum
computation. Hence, this scheme significantly simplifies the
experimental implementation of linear optical quantum computation
(LOQC)

A crucial requirement in the schemes of LOQC is the so-called
\emph{classical feedforwardability}, that is, it must be in
principle possible to detect when the gate has succeeded by
performing some appropriate measurement on ancilla photons
\cite{klm01,nielsen04}. This information can then be feedforward
for conditional future operations on the photonic qubits to
achieve efficient LOQC .

\begin{figure}
[ptb]
\begin{center}
\includegraphics[
height=2.9464in, width=2.4267in
]%
{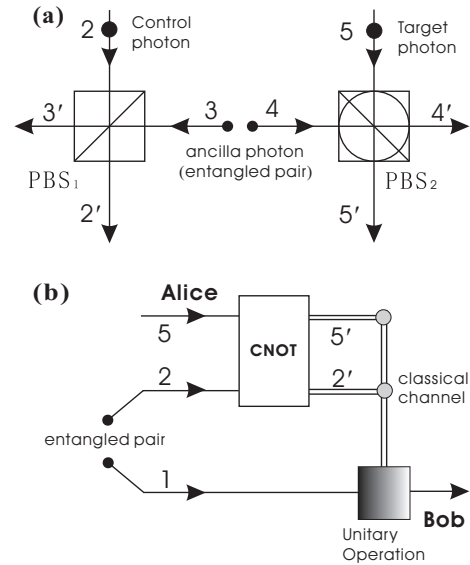}%
\caption{(a) A nondestructive CNOT gate constructed by polarizing
beam splitters (PBS), half-wave plates (HWP) and an ancilla
entangled photon pair $|\Psi^{-}\rangle_{34}$ \cite{pittman01}.
Conditioned on detecting a $|-\rangle$ photon in mode $3^{\prime}$
and a $|H\rangle$ photon in mode $4^{\prime}$ one can implement
the CNOT operation between the photons 2 and 5. (b) Quantum
circuit for quantum teleportation based on a CNOT gate
\cite{gotteman}. By using the CNOT operation, Alice can
discriminate the four orthogonal Bell state simultaneously such
that a complete teleportation can be
achieved.}%
\end{center}
\end{figure}

Recently destructive CNOT operations have been realized using
linear optical elements
\cite{pittman02,sanaka02,pittman03,white03}. However, as they
necessarily destroy the output state such logic operations are not
classically feed-forwardable and have little practical
significance. Fortunately, it has been suggested \cite{pittman01}
that a destructive CNOT gate together with the quantum parity
check can be combined with a pair of entangled photons to
implement a nondestructive (conventional) CNOT gate that satisfies
the feed-forwardability criterion.

In this paper, we present for the first time a proof-of-principle
demonstration of a non-destructive CNOT gate for two independent
photons, realizing the proposal of Pittman, Jacobs, and Franson
\cite{pittman01}. The quality of such a CNOT gate is further
demonstrated by discriminating all the four Bell-states in an
experiment on quantum teleportation.

In our experiment, we consider qubits implemented as the
polarization states of photons. We define the horizontal
polarization state $|H\rangle$ as logic 1, and the vertical one
$|V\rangle$ as logic 0. As shown in Fig. 1a, one can achieve the
desired nondestructive CNOT gate for photons 2 and 5 by performing
a quantum parity check on photons 2 and 3 and a destructive CNOT
operation on photons 4 and 5, where photons 3 and 4 are in the
state $|\Psi^{-}\rangle _{34}$, which is one of the four Bell
states
\begin{equation}%
\begin{array}
[c]{c}%
|\Phi^{\pm}\rangle_{ij}=\frac{1}{\sqrt{2}}(|H\rangle_{i}|H\rangle_{j}%
\pm|V\rangle_{i}|V\rangle_{j}),\\
|\Psi^{\pm}\rangle_{ij}=\frac{1}{\sqrt{2}}(|H\rangle_{i}|V\rangle_{j}%
\pm|V\rangle_{i}|H\rangle_{j}).
\end{array}
\label{epr}%
\end{equation}
Here $i$ and $j$ index the spatial mode of the photons. Then,
according to ref. \cite{pittman01} the nondestructive CNOT gate
for photons 2 and 5 can be accomplished conditioned on detecting a
$|-\rangle$ photon in mode $3^{\prime }$ and a $|H\rangle$ photon
in mode $4^{\prime}$. The logic table of the CNOT operation is
given by $|V\rangle|V\rangle\rightarrow|V\rangle|V\rangle$,
$|V\rangle|H\rangle\rightarrow|V\rangle|H\rangle$,
$|H\rangle|V\rangle \rightarrow|H\rangle|H\rangle$ and
$|H\rangle|H\rangle\rightarrow |H\rangle|V\rangle$.

One immediate application of the proposed CNOT gate is that it can
be used to generate entanglement between the control qubit and
target qubit \cite{barenco95}. For example, by setting the control
bit to be in the state $|-\rangle$ and the target qubit in the
state $\left\vert H\right\rangle $, one can utilize the
non-destructive CNOT gate to prepare photons 2 and 5 in the
entangled state $|\Psi^{-}\rangle_{25}$.

Another important application is that the non-destructive CNOT
gate can be used to simultaneously identify all the four Bell
states \cite{barenco95} in a quantum teleportation protocol
\cite{bennett93}. For example, suppose photon 5, which Alice wants
to teleport to Bob, is in an unknown polarization state
$|\Phi\rangle_{5}=\alpha\left\vert H\right\rangle
_{5}+\beta\left\vert V\right\rangle _{5}$, and the pair of the
photons $1$ and $2$ shared by Alice and Bob is in the entangled
state $\left\vert \Psi^{-}\right\rangle _{12}$ (Fig. 1b). It is
necessary to discriminate the four Bell-states of photons 2 and 5,
$\left\vert \Psi^{\pm}\right\rangle _{25}$ and $\left\vert
\Phi^{\pm }\right\rangle _{25}$ in order to realize the complete
quantum teleportation \cite{bennett93}.

Under the CNOT operation the four Bell-states of photons 2 and 5
will evolve into one of the four orthogonal separable states
\begin{equation}%
\begin{array}
[c]{c}%
\left\vert \Psi^{\pm}\right\rangle
_{25}\longrightarrow|\pm\rangle_{2^{\prime
}}\left\vert H\right\rangle _{5^{\prime}},\\
\left\vert \Phi^{\pm}\right\rangle
_{25}\longrightarrow|\pm\rangle_{2^{\prime }}\left\vert
V\right\rangle _{5^{\prime}}.
\end{array}
\end{equation}

Therefore, the crucial Bell-state measurement can be accomplished
probabilistically by applying a nondestructive CNOT operation and
performing a subsequent polarization analysis on photons 2 and 5.
Depending on Alice's measurement results, Bob can then perform a
unitary transformation, independent of $|\Phi\rangle_{5}$, on
photon 1 to convert its state into the
initial state of photon 5.%

\begin{figure}
[ptb]
\begin{center}
\includegraphics[
height=3.563in, width=3.435in
]%
{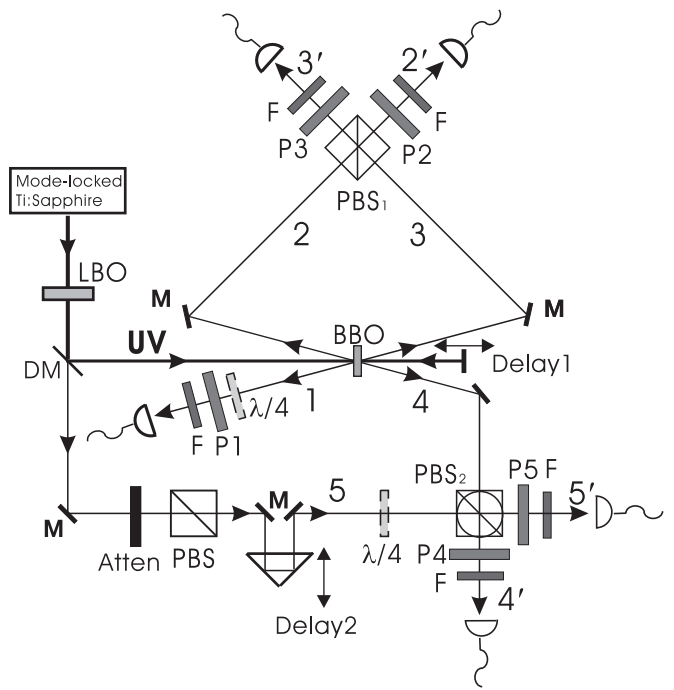}%
\caption{The experimental setup. Femtosecond laser pulses
($\approx$ 200fs, 76MHz, 788nm) first pass through a doubler LBO
crystal (LiB$_{3}$O$_{5}$), then the mixed ultraviolet (UV) and
infrared components were further separated by a dichroic
beamsplitter (DM). The reflected UV pulses passing through the BBO
crystal twice generate two pairs of entangled photons, one in
modes 1-2 and the other in modes 3-4, where both pairs are
prepared in the $|\Psi ^{-}\rangle$ state \cite{kwiat95,dik01}.
The observed coincidence rate of entangled photon pairs is about
$2.4\times10^{4}/$s. The photons in modes 3 and 4 are used as the
ancilla pair while the photon in mode 2 as the control qubit. The
arbitrary polarization state of the control photon can be readily
prepared by performing a polarization projection measurement on
photon 1. In the experiment, in order to prepare the target qubit,
i.e. photon 5, the transmitted near-infrared pulses are attenuated
by penetrating a ultra-fast laser output coupler and two
polarizers (i.e. the Atten) to a weak coherent beam such that
there is only a very small probability of containing a single
photon for each pulse (about 0.05 photon per pulse). The quarter
wave plate ($\lambda/4$) in modes 1 and 5 are used to prepare and
analyze the input and output circular polarized states,
respectively. The five polarizers P1, P2,
... P5 are used for polarization analysis.}%
\end{center}
\end{figure}

A schematic of the experimental setup used to demonstrate both the
CNOT gate and the identification of the four Bell states required
for quantum teleportation is shown in Fig. 2. To realize the CNOT
gate, it is necessary to overlap photons 2 and 3 at the PBS$_{1}$
and photons 4 and 5 at the PBS$_{2}$. In the experiment the
PBS$_{2}$, i.e. the desired 45-degree oriented polarizing
beamsplitter, is accomplished by inserting one half-wave plate
(HWP) in each of the two inputs and two outputs of an ordinary
polarizing beamsplitter. Note that, all the four HWP were oriented
at $22.5^{0}$ with respect to the horizontal direction, which
corresponds to a 45$^{0}$ polarization rotation. The good temporal
overlap was achieved by adjusting the two delay mirrors, Delay 1
and Delay 2. Experimentally, we first adjust the position of Delay
1 such that photons 2 and 3 arrive at the PBS$_{1}$
simultaneously, and then adjust the position of Delay 2 to achieve
the temporal overlap of photons 4 and 5 at the PBS$_{2}$.
Furthermore, the use of narrow-band interference filters (F) with
$\Delta\lambda_{FWHM}=3$ nm for all five photons makes the photons
at the same PBS indistinguishable \cite{marek}. The temporal
overlap between photons 2 and 3 was verified by observing a
four-particle interference visibility of $0.82$ among photons 1,
2, 3 and 4 by removing the PBS$_{2}$; and the temporal overlap of
photons 4 and 5 was verified by observing a three-particle
interference visibility of $0.68$ among photons 3, 4 and 5 by
removing the PBS$_{1}$ \cite{ghz1,ghz2}.

To experimentally demonstrate that the CNOT gate has been
successfully implemented, we first prepare the input control and
target qubits in the following specific states
$|H\rangle_{2}|H\rangle_{5}$, $|H\rangle
_{2}|V\rangle_{5}$, $|V\rangle_{2}|H\rangle_{5}$ and $|V\rangle_{2}%
|V\rangle_{5}$. If the CNOT gate works properly, then conditioned
on detecting a $|-\rangle$ polarized photon in mode $3^{\prime}$
and a $|H\rangle$ polarized photon in mode $4^{\prime}$ the two
qubits in modes $2^{\prime}$ and
$5^{\prime}$ would be, respectively, in the states $|H\rangle_{2^{\prime}%
}|V\rangle_{5^{\prime}}$,
$|H\rangle_{2^{\prime}}|H\rangle_{5^{\prime}}$,
$|V\rangle_{2^{\prime}}|H\rangle_{5^{\prime}}$ and $|V\rangle_{2^{\prime}%
}|V\rangle_{5^{\prime}}$. After the non-destructive CNOT operation
the output components corresponding to the above specific input
states, which were measured in the $H/V$ basis, are shown in Fig.
3a, respectively. The experimental fidelity of achieving the CNOT
logic table is estimated to be
$0.78\pm0.05$.%

\begin{figure}
[ptb]
\begin{center}
\includegraphics[
height=4.1987in, width=2.6654in
]%
{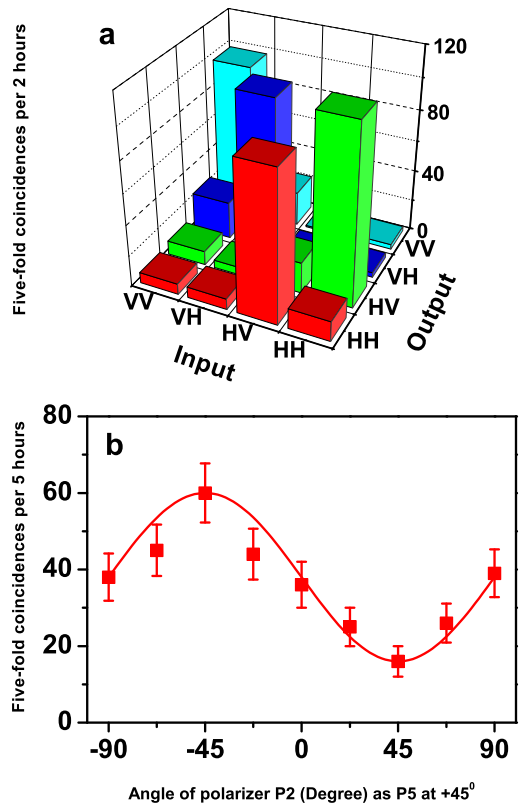}%
\caption{(a) Experimental results for the CNOT operation in the
$H/V$ basis. When the input control and target qubits are in the
following specific states $|H\rangle_{2}|H\rangle_{5}$,
$|H\rangle_{2}|V\rangle_{5}$, $|V\rangle _{2}|H\rangle_{5}$ and
$|V\rangle_{2}|V\rangle_{5}$, the output two qubits in modes
$2^{\prime}$ and $5^{\prime}$ would be, respectively, in the
states
$|H\rangle_{2^{\prime}}|V\rangle_{5^{\prime}}$, $|H\rangle_{2^{\prime}%
}|H\rangle_{5^{\prime}}$,
$|V\rangle_{2^{\prime}}|H\rangle_{5^{\prime}}$ and
$|V\rangle_{2^{\prime}}|V\rangle_{5^{\prime}}$.(b) Experimental
results for the CNOT operation when the control qubit was in the
state $\frac{1}{\sqrt{2}}\left( \left\vert H\right\rangle
-\left\vert V\right\rangle \right)  $ and the target qubit in the
state $\left\vert H\right\rangle $. After the CNOT operation the
two output qubits would be in the Bell state $\left\vert
\Psi^{-}\right\rangle $. We perform a conditional measurement of
polarization as a function of the orientation of polarizer P2 when
polarizer P5 was fixed at $+45^{0}$.}%
\end{center}
\end{figure}

Second, to show the CNOT gate also works for an arbitrary
superposition of the
control qubit, we now prepare the control qubit in the state $\frac{1}%
{\sqrt{2}}\left(  \left\vert H\right\rangle -\left\vert
V\right\rangle \right)  $ and the target qubit in the state
$\left\vert H\right\rangle $ to entangle these two independent
photons. Then, after the CNOT operation the two output qubits
would be in the state $\frac{1}{\sqrt{2}}\left(  \left\vert
H\right\rangle \left\vert V\right\rangle -\left\vert
V\right\rangle \left\vert
H\right\rangle \right)  $, i.e. the Bell state $\left\vert \Psi^{-}%
\right\rangle $. To verify the expected Bell state has implemented
successfully, we first measured the four possible polarization
combinations of the control and target qubits in the $H/V$ basis.
The signal-to-noise ratio between the desired
($|H\rangle|V\rangle$ and $|V\rangle|H\rangle$) and unwanted
($|H\rangle|H\rangle$ and $|V\rangle|V\rangle$) were measured to
be $4.2:1$. This confirms that the $|H\rangle|V\rangle$ and
$|V\rangle|H\rangle$ terms are the dominant components.
Furthermore, to prove the two terms are indeed in a coherent
superposition, we also perform a conditional coincidence
measurement as a function of the orientation of polarizer P2 as
polarizer P5 was fixed at $+45^{0}$. As shown in Fig. 3b, the
experimental results of the polarization correlation exhibit an
interference fringe with a visibility of $0.58\pm0.09$, which is
in consistent with the prediction of the interference fringe
for the Bell state $\left\vert \Psi^{-}\right\rangle $.%

\begin{figure}
[ptb]
\begin{center}
\includegraphics[
height=3.8908in, width=2.7916in
]%
{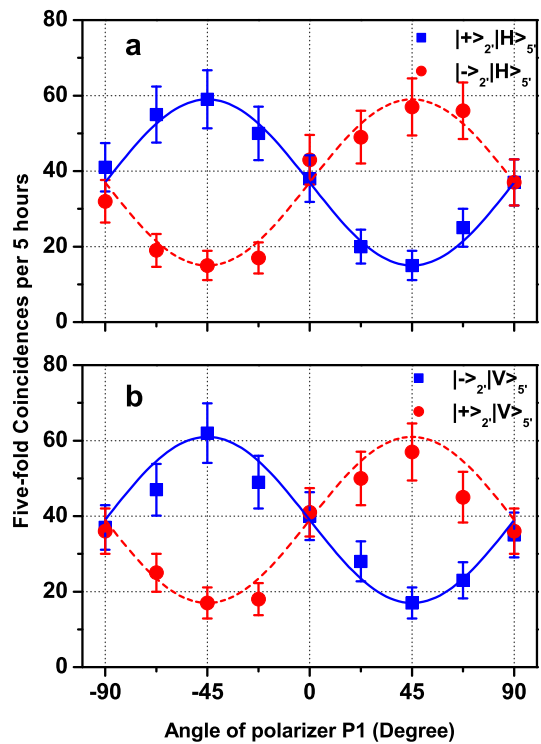}%
\caption{The experimental results for quantum teleportation with
complete Bell-state analysis. The data clearly confirm the
expected $\pi$ phase shift, hence demonstrating that the four
Bell-states have been identified successfully in the teleportation experiment.}%
\end{center}
\end{figure}

The CNOT gate can be used not only to entangle two independent
photons, it can also be used to disentangle two entangled photons.
To demonstrate the latter, let us now exploit the CNOT gate to
simultaneously discriminate all the four Bell-states in an quantum
teleportation experiment. As described in equation (2),
conditioned on detecting a $|-\rangle$ photon in mode $3^{\prime}$
and a $|H\rangle$ photon in mode $4^{\prime}$ the required joint
Bell-state measurement can be achieved by performing a
polarization measurement both on photon $2^{\prime}$ in the $+/-$
basis and on photon $5^{\prime}$ in the $H/V$ basis. For example,
registering a $|+\rangle_{2^{\prime}}\left\vert H\right\rangle
_{5^{\prime}}$ coincidence implies a projection onto the
Bell-state $|\Psi^{+}\rangle_{25}$. In this way, we can identify
all the four Bell-states. According to teleportation protocol
\cite{bennett93} it is obvious that, if the photons 2 and 5 are
measured to be in the state $|\Psi^{-}\rangle_{25}$, then photon 1
will be projected into the state $\alpha|H\rangle+\beta|V\rangle$;
if the photons 2 and 5 are measured to be in the state
$|\Psi^{+}\rangle_{25}$, then photon 1 will be left in the state
$\alpha|H\rangle-\beta|V\rangle$. In these two cases, the two
corresponding states of photon 1 would, in general, have a
relative phase shift of $\pi$
\cite{shih01}. Similarly, for the projections onto the state $|\Phi^{+}%
\rangle_{25}$ or $|\Phi^{-}\rangle_{25}$, photon 1 will be
correspondingly left in the state $\alpha|V\rangle-\beta|H\rangle$
or $\alpha|V\rangle +\beta|H\rangle$. Again, the two states of
photon 1 have a relative phase shift of $\pi$.

To experimentally verify the above analysis, we decided to
teleport the left-hand circular polarization state
$\frac{1}{\sqrt{2}}(|H\rangle -i|V\rangle)$ from photon 5 to
photon 1. The output circular polarization states of photon 1 are
analyzed by inserting a QWP and a polarizer in front of the
detector D1. As shown in Fig. 4a and b, the five-fold coincidences
are recorded as polarizer 1 was rotated. The experimental results
are consistent with the prediction of the $\pi$ phase shift, hence
confirming that the four Bell-states have been successfully
discriminated. Following the same definition of ref.
\cite{shih01}, our experimental average fidelity of teleportation
was estimated to be $F\approx0.79\pm0.05$, which clearly surpasses
the classical limit of $2/3$. Note that throughout the whole
experiment the fidelities were obtained without performing any
background subtraction.

Finally, we emphasize that in the above teleportation experiment
we only verified the corresponding relation between the initial
state of photon 5 and the final state of photon 1 after the
Bell-state analysis is complete, but did not perform a conditional
operation on the photon 1 to convert its final state into the
original state of photon 5. We plan to address this challenging
task in a forthcoming experiment.

In summary, we have for the first time experimentally demonstrated
a probabilistic nondestructive CNOT gate for two independent
photons using only linear optics. Furthermore, we demonstrated
that such a device can be used not only to entangle two
independent photons, but also to discriminate all the four
Bell-states for quantum teleportation. We believe that the methods
developed for this experiment would have various novel
applications in quantum information processing with linear optics.

This work was supported by the NSF of China, the CAS and the
National Fundamental Research Program (under Grant No.
2001CB309303), and the Alexander von Humboldt Foundation.

\end{document}